\documentclass[11pt,a4paper]{article}
\usepackage[left=20mm, top=20mm, right=20mm, bottom=20mm, nohead=true, nofoot=true]{geometry}
\usepackage[utf8]{inputenc}
\usepackage{authblk}
\usepackage{indentfirst}
\usepackage{misccorr}
\usepackage{graphicx}
\usepackage{amsmath}
\usepackage{amssymb}
\usepackage{multicol}
\usepackage{bm}
\usepackage{longtable}
\usepackage{pdflscape}
\usepackage{epstopdf}
\usepackage{multirow}
\usepackage{adjustbox}
\usepackage{amsfonts}
\usepackage{ifthen}
\usepackage{fancyhdr}
\usepackage{setspace}
\usepackage{xcolor}
\usepackage[round,authoryear,comma,sort&compress,numbers]{natbib}
\usepackage[toc,title,page]{appendix}
\usepackage[hidelinks]{hyperref}
\usepackage{url}

\hypersetup{
    colorlinks=true,
    linkcolor=green,
    citecolor=blue,
    filecolor=magenta,
    urlcolor=cyan,
    pdftitle={Sharelatex Example},
    pdfpagemode=FullScreen,
}

\fancypagestyle{plain}{\chead{\texttt{\textcolor{brown}{Communications of BAO, Vol. ?, Issue ?, 20??, pp. ???-???}}}}
\title{\textbf{On the simultaneous generation of radio and soft X-ray emission by AXP 4U 0142+61}}
\author[1]{Osmanov Z.N.\thanks{z.osmanov@freeuni.eduge, Corresponding author}}
\affil[1]{\scriptsize School of Physics, Free University of Tbilisi, 0183, Tbilisi,
Georgia}
\affil[2]{\scriptsize E. Kharadze Georgian National Astrophysical Observatory, Abastumani, 0301, Georgia}

\begin{document}
\pagestyle{empty}
\newpage
\pagestyle{fancy}
\label{firstpage}
\date{}
\maketitle

\begin{abstract}
In the present paper we study the possibility of a simultaneous
generation of radio waves and soft $X$-rays by means of the
quasi-linear diffusion (QLD) in the anomalous pulsar AXP 4U 0142+61.
Considering the magnetosphere composed of the so-called beam
component and the plasma component respectively, we argue that the
frozen-in condition will inevitably lead to the generation of the
unstable cyclotron waves. These waves, via the QLD, will in turn
influence the particle distribution function, leading to certain
values of the pitch angles, thus to an efficient synchrotron
mechanism, producing soft $X$-ray photons. We show that for
physically reasonable parameters of magnetospheric plasmas, the QLD
can provide generation of radio waves in the following interval $40$
MHz-$111$ MHz connected to soft $X$-rays for the domain
$0.3$keV-$1.4$keV.
\end{abstract}

\section{Introduction}

Anomalous X-ray pulsars (AXPs) (young isolated neutron stars) since
their discovery \citep{axp2,axp1} deserve a great attention despite
a few number of known AXPs \citep{kaspi}. These objects are
intensively studied last several years, but their nature still
remains unknown. One of the interesting features of AXPs is their
long period of rotation, which in turn leads to very strong magnetic
fields exceeding the so-called Schwinger limit, $B_{cr}\approx
4.41\times 10^{13}$G. Therefore, they are called magnetars. AXPs
exhibit strong X-ray fluxes and a corresponding luminosity exceeds
the spin-down luminosity by many orders of magnitude. On the other
hand, despite some predictions, that magnetars must be dark in the
radio band \citep{bhard}, \citet{camilo} and \citet{malof} reported
the detection of radio pulsations from magnetar-type neutron stars.
In particular, \citet{camilo} observed the position of the anomalous
pulsar XTE J1810 - 197 at frequencies from $\nu = 1.4$GHz to $\nu =
49$GHz. It was shown that XTE J1810 - 197 emits bright, narrow,
highly linearly polarized radio pulses. \citet{malof}, based on two
high-sensitivity radio telescopes of the Pushchino Radio Astronomy
Observatory - the Large Phased Array and the DKR-1000, have reported
the detection of weak radio pulsed emission from the X-ray pulsar
AXP 4U 0142+61 at two low frequencies, $40$MHz and $111$MHz. It is
worth noting that this pulsar was monitored by the Westerbork
Synthesis Radio Telescope at a frequency $1380$ MHz. The
observations did not detect a source of radio emission (with $1380$
MHz) at the location of AXP 4U 0142+61.

In the present paper we focus on the anomalous $X$-ray pulsar 4U
0142+61, which exhibits radiation from the soft- to hard- $X$-rays,
\citep{gohler,hartog,enoto}. The aim of this work is to study the
possibility of a simultaneous generation of soft $X$-rays and radio
waves stimulated by the quasi-linear diffusion (QLD). For explaining
the radiation in the soft $X$-rays, we account for the synchrotron
emission process. But, since in the magnetospheres of magnetars.
magnetic fields are very strong, the corresponding energy loses are
efficient and for studying the synchrotron radiation one has to take
into account a certain mechanism balancing the dissipative factors.
This in turn leads to the one dimensional distribution function of
particles and as a result the synchrotron mechanism completely
vanishes. In this paper we rely on the pulsar emission model
developed by \citet{machus1,lomin}. According to this approach, in
the pulsar magnetospheres the cyclotron instability appears
\citep{kmm}, which during the quasi-linear stage, causes a diffusion
of particles along and across the magnetic field lines, leading to
the required balance.

This mechanism was applied to magnetars, pulsars and active galactic nuclei in
a series of papers:
\citep{magnetar,malmach,difus,difus1,difus3,difus4,difus5,ninoz}. One of
the interesting consequences of the QLD is the fact that it provides
a simultaneous generation of waves in two different emission bands:
relatively low energy- and high energy- domains. In particular, the
high energy radiation appears by means of the feedback of the
cyclotron waves on relativistic particles due to the diffusion, and
as a result, the pitch angles are arranged according to the
aforementioned balance. Therefore, during the QLD, the physical
system will be characterized by two radiation regimes: (a) the high
energy synchrotron mechanism and (b) a low energy emission process
provided by the cyclotron waves. In this context the recent
observations performed by the MAGIC Cherenkov telescope deserve a
great interest. In particular, \citet{magic} reported about the
discovery of the very high energy (VHE) pulsed emission ($>25$ GeV)
from the Crab pulsar and it has been shown that the VHE signals are
coincident with optical signals in a phase. For explaining the
origin of the coincidence, \citet{difus} have considered the
mechanism of the QLD applying it to the plasmas in the magnetosphere
of the Crab pulsar on the light cylinder (a hypothetical area where
the linear velocity of rigid rotation exactly equals the speed of
light) lengthscales. We have found that the QLD provides the
simultaneous generation of emission in different frequency bands. In
the later studies \citep{difus1,ninoz} the same problem was examined
in more detail.

In the present paper we consider the anomalous pulsar 4U 0142+61 to
investigate the role of the QLD in generation of the detected soft
$X$-rays and radio waves respectively. The paper is organized as
follows: In Section 2 we introduce the mechanism of the QLD, in
Sect. 3 we apply the method to AXP 4U 0142+61 and obtain results,
and in Sect. 4 we summarize them.

%%%%%%%%%%%%%%%%%%%%%%%%%%%%%%%%%%%%%%%%
\section{Main consideration} \label{sec:consid}
%%%%%%%%%%%%%%%%%%%%%%%%%%%%%%%%%%%%%%%%

We assume that the pulsar's magnetosphere is composed of the
so-called primary beam with the Lorentz factor, $\gamma_b$ and the
bulk component with the Lorentz factor, $\gamma_p$
\citep{difus,difus1,ninoz}. By \citet{kmm} it was shown that in the
pulsar magnetospheric plasmas, which satisfy the frozen-in
condition,  the anomalous Doppler effect induces resonance unstable
cyclotron waves
\begin{equation}\label{res}
\omega - k_{_{\|}}c-k_xu_x-\frac{\omega_B}{\gamma_{b}} = 0
\end{equation}
with the corresponding frequency \citep{malmach}
\begin{equation}\label{nu}
\nu \approx \frac{\omega_{_B}}{2\pi\delta\gamma_b},\;\;\;\;\; \delta
= \frac{\omega_p^2}{4\omega_{_B}^2\gamma_p^3},
\end{equation}
where $k_{_{\|}}$ is the longitudinal (along the magnetic field
lines) component of the wave vector, $u_x\approx
c^2\gamma_b/(\rho\omega_{_B})$ is the so-called curvature drift
velocity, $c$ is the speed of light, $\rho$ is the magnetic fields'
curvature radius, $k_x$ is the wave vector's component along the
drift,  $\omega_{_B}\equiv eB/mc$ is the cyclotron frequency,
$B\approx 2.35\times 10^{14}R_{st}^3/R^3$G is the magnetic induction
close to the star's surface, $R_{st}\approx 10^6$cm is the pulsar's
radius, $R$ is the distance from the pulsar's center, $e$ and $m$
are the electron's charge and the rest mass respectively, $\omega_p
\equiv \sqrt{4\pi n_pe^2/m}$ is the plasma frequency and $n_p$ is
the plasma number density.

For studying the development of the QLD, one should note that two
major forces control dissipation. When particles emit in the
synchrotron regime, they undergo the radiative reaction force ${\bf
F}$, having the following components \citep{landau}:
\begin{equation}\label{fs}
    F_{\perp}=-\alpha_{s}\frac{p_{\perp}}{p_{\parallel}}\left(1+\frac{p_{\perp}^{2}}{m^{2}c^{2}}\right),
    F_{\parallel}=-\frac{\alpha_{s}}{m^{2}c^{2}}p_{\perp}^{2},
\end{equation}
where $\alpha_{s}=2e^{2}\omega_{_B}^{2}/3c^{2}$ and $p_{\perp}$ and
$p_{\parallel}$ are the transversal (perpendicular to the magnetic
field lines) and longitudinal (along the magnetic field lines)
components of the momentum respectively.

In nonuniform magnetic field, electrons also experience a force
${\bf G}$, that is responsible for conservation of the adiabatic
invariant, $I = 3cp_{\perp}^2/2eB$. The corresponding components of
${\bf G}$ are given by \citep{landau}:
\begin{equation}\label{g}
G_{\perp} = -\frac{cp_{\perp}}{\rho},\;\;\;\;\;G_{_{\|}} =
\frac{cp_{\perp}^2}{\rho p_{\parallel}}.
\end{equation}

The wave excitation leads to a redistribution process of the
particles via the QLD, which is described by the following kinetic
equation \citep{machus1,malmach1}
\begin{eqnarray} \label{qld1}
    \frac{\partial\textit{f }}{\partial
    t}+\frac{1}{p_{\perp}}\frac{\partial}{\partial p_{\perp}}\left(p_{\perp}
    \left[F_{\perp}+G_{\perp}\right]\textit{f }\right)=\nonumber \\
    =\frac{1}{p_{\perp}}\frac{\partial}{\partial p_{\perp}}\left(p_{\perp}
D_{\perp,\perp}\frac{\partial\textit{f }}{\partial
p_{\perp}}\right),
\end{eqnarray}
where $\textit{f }$ is the distribution function of the zeroth
order, $D_{\perp,\perp}=D\delta |E_k|^2$, is the diffusion
coefficient, $|E_k|^2$, is the energy density per unit of wavelength
and $D=e^{2}/8c$ \citep{ninoz}. For estimating $|E_k|^2$, it is
natural to assume that half of the plasma energy density,
$mc^2n_b\gamma_b/2$ converts to the energy density of the waves
$|E_k|^2k$, then for $|E_k|^2$ we obtain
\begin{equation}\label{ek2}
|E_k|^2 = \frac{mc^3n_b\gamma_b} {4\pi\nu},
\end{equation}
where
\begin{equation}\label{nb}
n_b = \frac{B} {Pce},
\end{equation}
is the number density of the beam and $P\approx 8.7$s is the
rotation period of the pulsar.

By taking into account the relations $\psi\equiv
p_{\perp}/p_{\parallel}$, $p_{\parallel}=mc\gamma_b$, one can show
from Eqs. (\ref{fs},\ref{g}) that
\begin{equation}\label{gfpp}
\frac{F_{\perp}}{G_{\perp}}\approx 2.7\times
10^{-6}\times\left(\frac{B}{10^3G}\right)^2\times\left(\frac{\gamma_b}{10^7}
\right)\times\left(\frac{\psi}{10^{-5}rad}\right)^2,
\end{equation}
where $B$ is normalized to the value of the magnetic field in the
magnetosphere on the lengthscales, $\sim 10^{10}$cm. We see from
this ratio that for physically reasonable parameters, one can
neglect the transversal component of the radiation reaction force.
Therefore, Eq. (\ref{qld1}) reduces to
\begin{eqnarray} \label{qld}
    \frac{\partial\textit{f }}{\partial
    t}+\frac{1}{p_{\perp}}\frac{\partial}{\partial p_{\perp}}\left(p_{\perp}
    G_{\perp}\textit{f }\right)=\nonumber \\
    =\frac{1}{p_{\perp}}\frac{\partial}{\partial p_{\perp}}\left(p_{\perp}
D_{\perp,\perp}\frac{\partial\textit{f }}{\partial
p_{\perp}}\right).
\end{eqnarray}

As it is clear from Eq. (\ref{qld}), two major factors compete in
this "game". On the one hand, the force responsible for conservation
of the adiabatic invariant attempts to decrease the transversal
momentum (thus the pitch angle), whereas the diffusion process, by
means of the feedback of the cyclotron waves, attempts to increase
the transversal momentum. Dynamically this process saturates when
the aforementioned factors balance each other. Therefore, it is
natural to study the stationary regime, $\partial\textit{f }
/\partial t=0$ and examine a saturated state of the distribution
function. After imposing the condition $\partial\textit{f }
/\partial t=0$ on Eq. (\ref{qld}) one can straightforwardly solve it
\begin{equation}\label{f}
    \textit{f}(p_{\perp})=C exp\left(\int \frac{G_{\perp}}{D_{\perp,\perp}}
    dp_{\perp}\right)=Ce^{-\left(\frac{p_{\perp}}{p_{\perp_{0}}}\right)^{2}},
\end{equation}
where $C={\it const}$ and
\begin{equation}\label{p0}
     p_{\perp_{0}}\equiv\left(\frac{2\rho D_{\perp,\perp}}{c}\right)^{1/2}.
\end{equation}
Since $\textit{f}$ is a continuous function of the transversal
momentum, it is natural to examine an average value of it and
estimate the corresponding mean value of the pitch angle,
$\bar{\psi}\equiv \bar{p}_{\perp_{0}}/p_{\parallel}$,
\begin{equation}\label{pitch}
\bar{\psi}
 = \frac{1}{p_{\parallel}}\frac{\int_{0}^{\infty}p_{\perp} \textit{f}(p_{\perp})dp_{\perp}}{\int_{0}^{\infty}\textit{f}(p_{\perp})dp_{\perp}}
\approx \frac{1}{\sqrt{\pi}}\frac{p_{\perp_{0}}}{p_{\parallel}}.
\end{equation}
As the investigation shows, the QLD leads to a certain distribution
of particles with the pitch angles, which will inevitably result in
the synchrotron radiation mechanism with the following energy of
emitted photons \citep{Lightman}
\begin{equation}
\label{eps} \epsilon_{eV}\approx 1.2\times
10^{-8}B\gamma_b^2\sin\bar{\psi}.
\end{equation}
\begin{figure}
  \resizebox{\hsize}{!}{\includegraphics[angle=0]{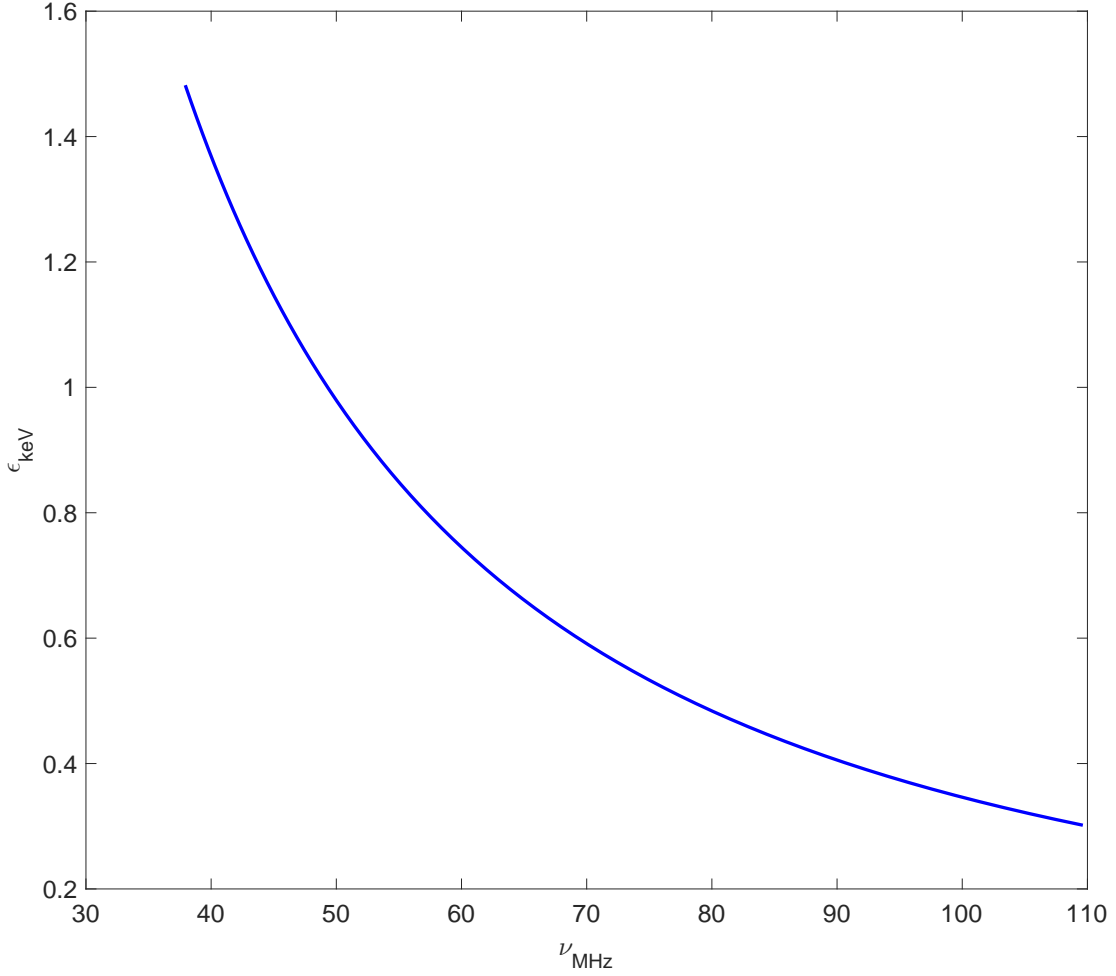}}
  \caption{Behaviour of $\epsilon_{keV}$ with respect to $\nu_{_{MHz}}$.
The set of parameters is: $P\approx 8.7$, $\gamma_p=2.77$,
$B_{st}\approx 1.3\times 10^{14}$G, $R_{st}\approx 10^6$cm,
$R=4.8\times 10^9$cm.}\label{fig1}
\end{figure}
\section{Results}\label{sec:results}

In this section we will apply the mechanism of the quasi-linear
diffusion to the anomalous pulsar 4U 0142+61 for studying the
possibility of simultaneous generation of radio waves and soft
$X$-rays. In the framework of the proposed model, the QLD is
provided by the feedback of the cyclotron waves. Let us consider
mildly relativistic particles of the plasma component with
$\gamma_p=3$ and the beam component with $\gamma_b = 10^7$. Then, by
taking into account that the energy is uniformly distributed,
$n_b\gamma_b\approx n_p\gamma_p$, one can reduce Eq.(\ref{nu})
\begin{equation}
\label{nu1} \nu\approx 1.9\times\left(\frac{\gamma_p}{3}\right)^4
\times\left(\frac{\gamma_b}{10^7}\right)^{-2}\times\left(\frac{R}{10^{10}cm}\right)^{-6}
MHz.
\end{equation}
As we see from this expression, the cyclotron frequency is very
sensitive to a location in the magnetosphere. One can
straightforwardly show that considering the following interval of
the beam Lorentz factors $(1-2)\times 10^7$ the best fit to
observations ($40$MeV, $111$MeV) is achieved by the parameters,
$\gamma_p\sim 2.77$, $R\sim 4.8\times 10^9$cm. Despite the mentioned
fact that these are resonance cyclotron waves, we see that the
corresponding frequency interval is relatively wide. The reason is
following: the resonance happens for given values of the Lorentz
factors - thus for a certain value of it, there is a certain value
of the radiation frequency. But relativistic particles are
distributed by their kinetic energy, which lies in a broad interval.
Therefore, in exciting waves all resonance particles (with broad
energy spectra) participate and the resulting frequencies will have
a relatively broad interval as well.

As a next step we would like to estimate how efficient is the
mentioned instability. \citet{kmm} have shown that for
$\gamma_b/(2\rho\omega_B)\ll\delta$ (which is the case) the
increment characterizing amplification of the cyclotron waves is
given by
\begin{equation}
\label{inc1} \Gamma=\frac{\omega_b^2}{2\nu\gamma_p},
\end{equation}
where $\omega_b \equiv \sqrt{4\pi n_be^2/m}$ is the plasma frequency
corresponding to the beam component. One can show that for the
aforementioned parameters, the value of the growth rate lies in the
following interval $\sim 10^{3}-10^{4}$s$^{-1}$. Therefore, the
corresponding timescale, $\tau\sim 1/\Gamma$, is of the order of
$\sim 10^{-4}-10^{-3}$s. On the other hand, we have seen that the
best fit to observations is achieved for the waves excited in the
location, $R\sim 5.9\times 10^9$cm. This means that plasmas stay
inside the magnetosphere for relatively long time. In particular,
the escape timescale, $t_{esc}\sim (R_{lc}-R)/c$ ($R_{lc}=cP/(2\pi)$
is the light cylinder radius) is of the order of $\sim 1$s. As we
see, the instability timescale is by many orders of magnitude
less than the escape timescale, which means that the process is extremely
efficient and physically feasible.

We have already explained that the cyclotron waves will influence
the particle distribution via diffusion (feedback mechanism) leading
to certain pitch angles (see Eqs. (\ref{pitch},\ref{eps})). In Fig.
\ref{fig1} we show the dependence of synchrotron photon energy on
the radio frequency. The set of parameters is: $P\approx 8.7$s,
$\gamma_p=2.77$, $B_{st}\approx 1.3\times 10^{14}$G, $R_{st}\approx
10^6$cm, $R=4.8\times 10^9$cm. It is clear from the plot that
$\epsilon_{keV}$ is a continuously decreasing function of radio
frequencies. This is direct consequence of Eqs.
(\ref{nu},\ref{ek2},\ref{p0},\ref{pitch},\ref{eps}). In particular,
according to Eq. (\ref{eps}) the photon energy behaves as to be
$\epsilon_{eV}\sim\gamma_b\overline{\psi}$, on the other hand, by
taking into account the relation $D_{\perp,\perp}=D\delta |E_k|^2$,
one can see from Eqs. (\ref{ek2},\ref{p0},\ref{pitch}) that
$\overline{\psi}\sim\gamma_b$, which by combining with Eq.
(\ref{eps}) leads to the following dependence
$\epsilon_{eV}\sim\gamma_b^3$. Therefore, more energetic particles
produce more energetic synchrotron photons, but since the cyclotron
frequency is a decreasing function of $\gamma_b$ (see Eq.
(\ref{nu1})), lower radio frequencies correspond to higher $X$-ray
photon energies.

As it is clear from the plot, the relativistic electrons with
Lorentz factors $(1-1.7)\times 10^7$, can lead to a simultaneous
generation of radio waves (from $40$ MHz to $111$ MHz) and soft
$X$-rays (from $0.3-1.4$ keV) respectively. According to the
proposed model, emission mechanisms are produced by plasmas inside
the magnetosphere of the anomalous pulsar 4U 0142+61, relatively far
as from the neutron star's surface, as from the light cylinder area,
$R\sim 4.8\times 10^9$cm.

\section{Summary}\label{sec:summary}

The main aspects of the present work can be summarized as follows:
\begin{enumerate}

      \item In this paper we examined the role of the quasi-linear diffusion
      in producing soft $X$-rays and radio emission in the magnetosphere of
      the anomalous pulsar 4U 0142+61.

      \item Considering the anomalous Doppler effect, which leads to
      the unstable cyclotron waves, we have studied the feedback of
      these waves on a distribution of relativistic particles.
      Solving the equation governing the QLD, the corresponding
      expression of the average value of the pitch angle is derived
      and analyzed for physically reasonable parameters. It has been
      found that the higher the the synchrotron photon energy,
      the lower the radio frequency.

      \item We have shown that the quasi-linear diffusion might
      provide a simultaneous generation of radio emission ($40$MHz-$111$MHz) and soft
      $X$-rays ($0.3$keV-$1.4$keV) in plasmas located on the
      distance $4.8\times 10^9$cm from the pulsar's center for
      appropriate parameters $\gamma_p=2.77$, $\gamma_b=(1-1.7)\times 10^7$.

      \end{enumerate}

The present investigation shows that the QLD is a mechanism that can
explain a simultaneous generation of the observationally evident
radio waves \citep{malof} and soft $X$-rays \citep{gohler}. The aim
of the present paper was to examine only one part of the problem,
although a complete study requires to investigate the spectral
pattern of emission as well. In the standard theory of the
synchrotron emission it is assumed that due to the chaotic character
of the magnetic field lines \citep{bekefi,ginz}, the pitch angles
lie in a broad interval (from $0$ to $\pi/2$). In our model the
distribution function of particles is strongly influenced by the
process of the QLD and as a result the pitch angles are restricted
by the balance of dissipative and diffusive factors. This will
inevitably lead to a spectral pattern, different from that of
\citet{bekefi,ginz}. Therefore, we will investigate this problem in
future studies.

\section*{\small Acknowledgements}
\scriptsize{The ComBAO would like to the thank the dedicated researchers who are publishing with the ComBAO.}

%\bibliographystyle{ComBAO}
%\nocite{*}
%\bibliography{references}

\end{document}